\documentclass[cite]{epl}

\def\mgb2{MgB$_2$}
\def\mgbcx{MgB$_{2-x}$C$_x$}
\def\mgbc2{MgB$_{1.98}$C$_{0.02}$}
\def\mgbc4{MgB$_{1.96}$C$_{0.04}$}
\def\mgbc6{MgB$_{1.94}$C$_{0.06}$}
\def\musr{$\mu^+$SR}

\title{\musr\ study of carbon-doped \chem{MgB_2} superconductors}
\author{K. Papagelis\inst{1}, J. Arvanitidis\inst{1}, K. Prassides\inst{1}, A. Schenck\inst{2},\\
T. Takenobu\inst{3} and Y. Iwasa\inst{3}}
\institute{
  \inst{1} School of Chemistry, Physics and Environmental Science,
University of Sussex, \\
Brighton BN1 9QJ, UK\\
  \inst{2} Institute for Particle Physics, ETH Zurich,
 CH-5232 Villigen PSI, Switzerland \\
  \inst{3} Institute for Materials Research, Tohoku University, Sendai 980-8577 and \\ 
 CREST, Japan Science and Technology Corporation, Kawaguchi 332-0012, Japan \\
}

\shortauthor{K. Papagelis et al.}


\begin{document}

\maketitle

\begin{abstract}
The evolution of the superconducting properties of the carbon-doped \mgb2 superconductors, 
\mgbcx\ ($x$= 0.02, 0.04, 0.06) have been investigated by the transverse-field muon spin rotation (TF-\musr) technique. 
The low-temperature depolarisation rate, $\sigma$(0) at 0.6 T which is proportional 
to the second moment of the field distribution of the vortex lattice decreases monotonically with increasing 
electron doping and decreasing $T_c$. In addition, the temperature dependence of $\sigma$($T$) has been analysed 
in terms of a two-gap model. 
The size of the two superconducting gaps decreases linearly as the carbon content increases. 
\end{abstract}

\section{Introduction}
The recent discovery of superconductivity at $\sim$40 K in \mgb2~\cite{Nagamatsu} has opened new prospects for the 
understanding of the microscopic origin of high-$T_c$ superconductivity. In contrast to the complex 
crystal structures, multicomponent nature and complications due to magnetism and strong electron-electron 
correlations of the high-$T_c$ cuprates, \mgb2\ has a simple hexagonal crystal structure (AlB$_2$-type) 
comprising close-packed Mg$^{2+}$ layers alternating with graphite-like boron layers and $sp$ 
electrons involved in the superconducting process.
Several attempts~\cite{Slusky} have been made to 
introduce dopants in the Mg layers in order to explore the relationship between
$T_c$, crystal structure and doping level. 
The reported experiments invariably lead to a decrease in $T_c$. Successful substitution at the B sites can 
be also achieved by carbon doping~\cite{Takenobu}, which results in a significant contraction of the $a$ lattice 
parameter, but affects little the interlayer separation. 
$T_c$ also decreases with increasing doping.
Experimental studies of the superconducting properties of \mgb2 show deviations from those 
calculated with the standard BCS theory~\cite{Budko,Bouquet1}. Various experiments, including scanning 
tunneling microscopy (STM)~\cite{Giubileo}, point-contact spectroscopy~\cite{Szabo}, specific heat 
measurements~\cite{Wang}, optical~\cite{Kuzmenko} and Raman spectroscopy~\cite{Chen} suggested the existence of a secondary 
superconducting gap, implying that the simple one-band approach for \mgb2\ superconductivity must be extended. 
On the basis of the electronic structure, the existence of multiple gaps has been invoked 
in order to explain the magnitude of $T_c$ in \mgb2 \cite{Liu}. In addition, solution of the 
full Eliashberg equations at low temperature yields different gap values for the different parts of the Fermi 
surface ($\sim1.8$ meV for 3D sheets and $\sim6.8$ meV for 2D sheets)~\cite{Choi}. Although there is much 
support for the applicability of the multiband description to \mgb2, there is still some debate in the 
literature, in particular since some tunneling and NMR
measurements show only a single gap~\cite{Gonnelli}. 

\musr\ measurements have also made significant contributions to the understanding of the superconducting 
properties of \mgb2, mainly through the precise determination of the penetration depth, $\lambda$ from the 
zero-temperature extrapolated value of the $\mu^+$ spin depolarisation
rate, $\sigma (0)$. Early experiments~\cite{Panagopoulos} 
proposed that the low temperature magnetic penetration depth of \mgb2\ shows a quadratic 
temperature dependence and were interpreted in terms of unconventional superconductivity with an energy gap that has 
nodes at certain points in the $k$ space. However, a systematic study by Niedermayer {\it et al.}~\cite{Niedermayer} 
reported that 
the temperature evolution of the depolarisation rate, $\sigma$ in polycrystalline \mgb2 could be well interpreted in 
terms of a two-gap model.

In the present paper we employ the transverse-field (TF) variant 
of the \musr\ technique 
to characterise the superconducting properties of carbon-doped \mgb2\  
systems (\mgbcx, $x$= 0.02, 0.04, 0.06). Analysis of the experimental 
data in terms of a two-gap model reveals the effect of electron 
doping on the superconducting gaps and the anisotropic properties of the compounds.

\section{Experimental}
Polycrystalline \mgbcx\ ($x$= 0.02, 0.04, 0.06) samples were prepared by reaction of Mg, amorphous B and carbon black 
at 900$^{\circ}$C for 2 h, as described elsewhere~\cite{Takenobu}.
The \musr\ measurements were performed with the GPS spectrometer on the $\pi$M3 muon beamline at the Paul 
Scherrer Institute (PSI), Switzerland. 
Pressed sample pellets were attached with low-temperature varnish on a Ag sample-holder placed on the stick
of a He continuous flow cryostat operating down to 1.8 K. After cooling the sample 
in an external field, $H_{ext}$ to temperatures below $T_c$ in order to induce a homogenous flux line lattice, 
positive muons (100\% spin-polarised) with their initial spin polarisation transverse to the external field 
were implanted in the solid sample. The implanted muons come to rest at
an interstitial site and act as highly sensitive local magnetic probes. In the presence of local fields, 
$B_{loc}$, the $\mu^+$ spin undergoes Larmor precession with frequency, 
$\omega_{\mu}=\gamma_{\mu}B_{loc}$, where $\gamma_{\mu}/2\pi$=13.553 kHz/G is the muon gyromagnetic ratio. 
The time evolution of the $\mu^+$ spin polarisation, $P_{\mu}(t)$ is measured by monitoring the positrons 
preferentially emitted along the $\mu^+$ spin direction at the instant of muon decay. For type II 
superconductors, $P_{\mu}(t)$ is an oscillatory function with decreasing amplitude and the damping of the 
$\mu^+$ precession signal provides a measure of the inhomogeneity of the magnetic field, $\Delta B$ in the vortex 
state and hence of the magnetic penetration depth, $\lambda$.

\section{Results and Discussion}
For polycrystalline samples in the vortex state, the TF $\mu^+$ spin polarisation function is approximately Gaussian 
($P_{\mu}(t)\sim\exp{(-\frac{1}{2}\sigma^2t^2)}$) and the depolarisation rate, $\sigma$ 
is proportional to the second moment of the field distribution ($\sigma\sim<$$\Delta$$B^2$$>^{1 \over 2}$). 
In the case of anisotropic ($\lambda_c / \lambda_{ab} >>$ 4) 
type II superconductors, $\sigma$ is related to the in-plane 
magnetic penetration 
depth, $\lambda_{ab}$ via the relation~\cite{Barfold}:
$\sigma[\mu s^{-1}]=7.086\times10^4~\lambda_{ab}^{-2}[nm^{-2}]$. 
The anisotropy of the upper critical field in single crystalline
\mgb2\ was reported as 6 (at 5 K) \cite{Angst} or 4.3 \cite{Takahashi},
while for polycrystalline samples and thin films, the reported values
span a wide range.
The above equation holds in the London limit 
($\kappa=(\lambda/\xi)>>1$) and in the absence of pinning-induced distortions of the vortex lattice. 
The former restriction is valid for \mgb2\ ($\xi_{ab}\approx$ 7 nm~\cite{DeLima}, $\lambda_{ab}\approx$ 95 nm
\cite{Niedermayer}), 
while the latter is addressed below.
\begin{figure}
\onefigure[scale=0.38]{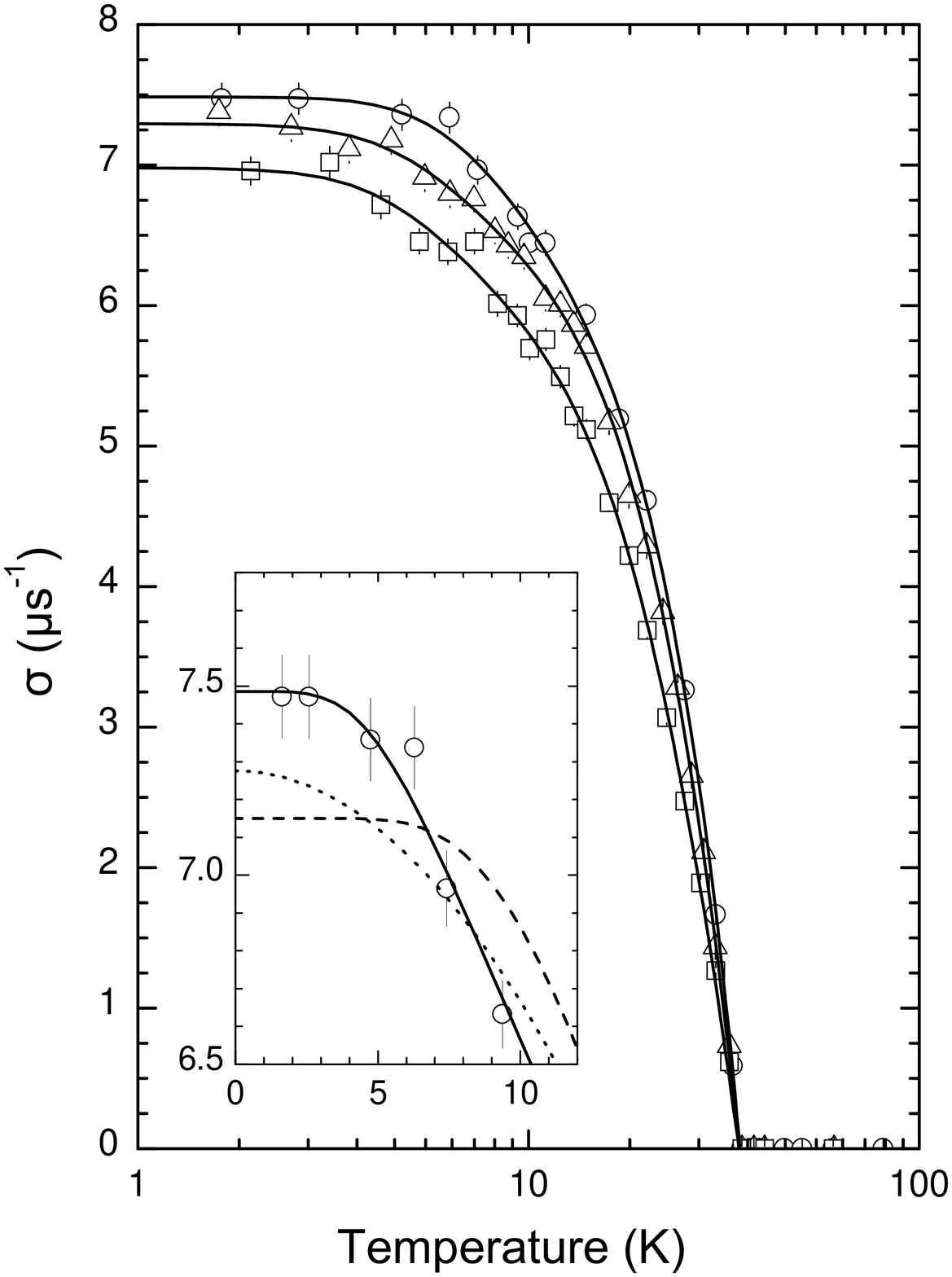}
\caption{Temperature dependence of the $\mu^+$ spin depolarisation rate, $\sigma$ 
at $H_{ext}$= 0.6 T for \mgbcx\ ($x$= 0.02, 0.04, 0.06). 
The open circles correspond to MgB$_{1.98}$C$_{0.02}$, 
the triangles to MgB$_{1.96}$C$_{0.04}$ and the squares to MgB$_{1.94}$C$_{0.06}$. 
The solid lines are fits to the two-gap model. The inset shows the 
fit of the experimental data for MgB$_{1.98}$C$_{0.02}$, assuming an isotropic single gap model (dashed line) 
or a $T^2$-law (dotted line).}
\label{f.1}
\end{figure}
The London model predicts that the second moment of the magnetic field distribution of a perfect vortex 
lattice should be independent of $H_{ext}$ for $\lambda > L$ where $L$
is the distance between vortices. 
In order to check the influence of pinning on the depolarisation 
rate, $\sigma$, we have measured its field dependence at 5 K. Our results 
show that $\sigma(H)$ 
increases almost linearly up to 50 mT, displays a peak at $\sim$80 mT and then decreases rapidly, reaching a plateau at 
higher fields ($\sim 0.6$ T) in excellent agreement with the reported data 
for \mgb2~\cite{Niedermayer}. 
We thus performed our measurements at an external 
field of 0.6 T in order to ensure formation of an ideal vortex lattice and avoid 
underestimation of $\lambda (T)$.

Fig.~\ref{f.1} presents the extracted temperature dependence of the TF-\musr\ depolarisation rate 
at $H_{ext}$= 0.6 T for the \mgbcx\ ($x$= 0.02, 0.04, 0.06) samples. The values of $\sigma (T)$ are derived 
after subtraction of the normal state temperature-independent depolarisation rate, $\sigma_{back}$
($\sigma^2 (T)=\sigma_{meas}^2 (T)-\sigma_{back}^2$). In the case of MgB$_{1.98}$C$_{0.02}$, $\sigma (T)$ 
increases monotonically as the temperature decreases below $\sim$35 K and reaches a plateau
at $T\leq$ 6 K, remaining almost constant at lower temperatures. Analogous behaviour 
is observed for the other two compositions, except that as the doping level increases, the low-temperature
plateau smooths out and the extrapolated value, $\sigma(0)$ shifts to lower values.

The experimental $\sigma(T)$ dependence is reproduced well by means of a two-gap model, while attempts to fit 
the experimental data with a $T^2$-law or an isotropic one-gap model led to unsatisfactory
results (inset in fig.~\ref{f.1}). 
The two-gap model is based on the existence of two discrete superconducting gaps, $\Delta_1$ and $\Delta_2$, at $T$= 0 K, 
both closing at $T_c$ and each associated with a different energy band. By assuming that the coupling between the 
two bands (i.e. due to impurity or phonon scattering) is sufficiently weak ($vide~infra$), 
$\sigma(T)$ can be expressed as~\cite{Wang,Bouquet1,Niedermayer}: 
\begin{equation}
\label{e.1}
\sigma(T)=\sigma(0)-(\gamma_1/\gamma)~\delta\sigma(\Delta_1,T)-(\gamma_2/\gamma)~\delta\sigma(\Delta_2,T)
\end{equation}
where
\begin{equation}
\label{e.2}
\delta\sigma(\Delta,T)=\frac{2\sigma(0)}{k_BT}\int_{0}^{\infty}f(\varepsilon,T) [1-f(\varepsilon,T)]d\varepsilon
\end{equation}
$f(\varepsilon,T)$ is the Fermi distribution of quasiparticles and $\varepsilon$ the energy of the normal 
electrons relative to the Fermi energy:
\begin{equation}
\label{e.3}
f(\varepsilon,T)=(1+\exp(\sqrt{\varepsilon^2+\Delta(T)^2}/k_BT))^{-1}
\end{equation}
Each band is characterised by a partial Sommerfeld constant, $\gamma_i$ 
($\gamma_1+\gamma_2=\gamma$, where $\gamma$ is the total Sommerfeld constant). As 
the Sommerfeld constant is proportional to the density-of-states at the Fermi level, the ratios $\gamma_i$/$\gamma$ 
determine the partial $N_i$($\epsilon _F$) for the two bands. 
The temperature dependence of the band gaps is taken 
from BCS theory, i.e. $\Delta(t)=\Delta(0)\delta(t)$ where $\delta(t)$ is the normalised BCS gap 
at the reduced temperature, $t=T/T_c$~\cite{Muhl}.
\begin{figure}
\onefigure[scale=0.35]{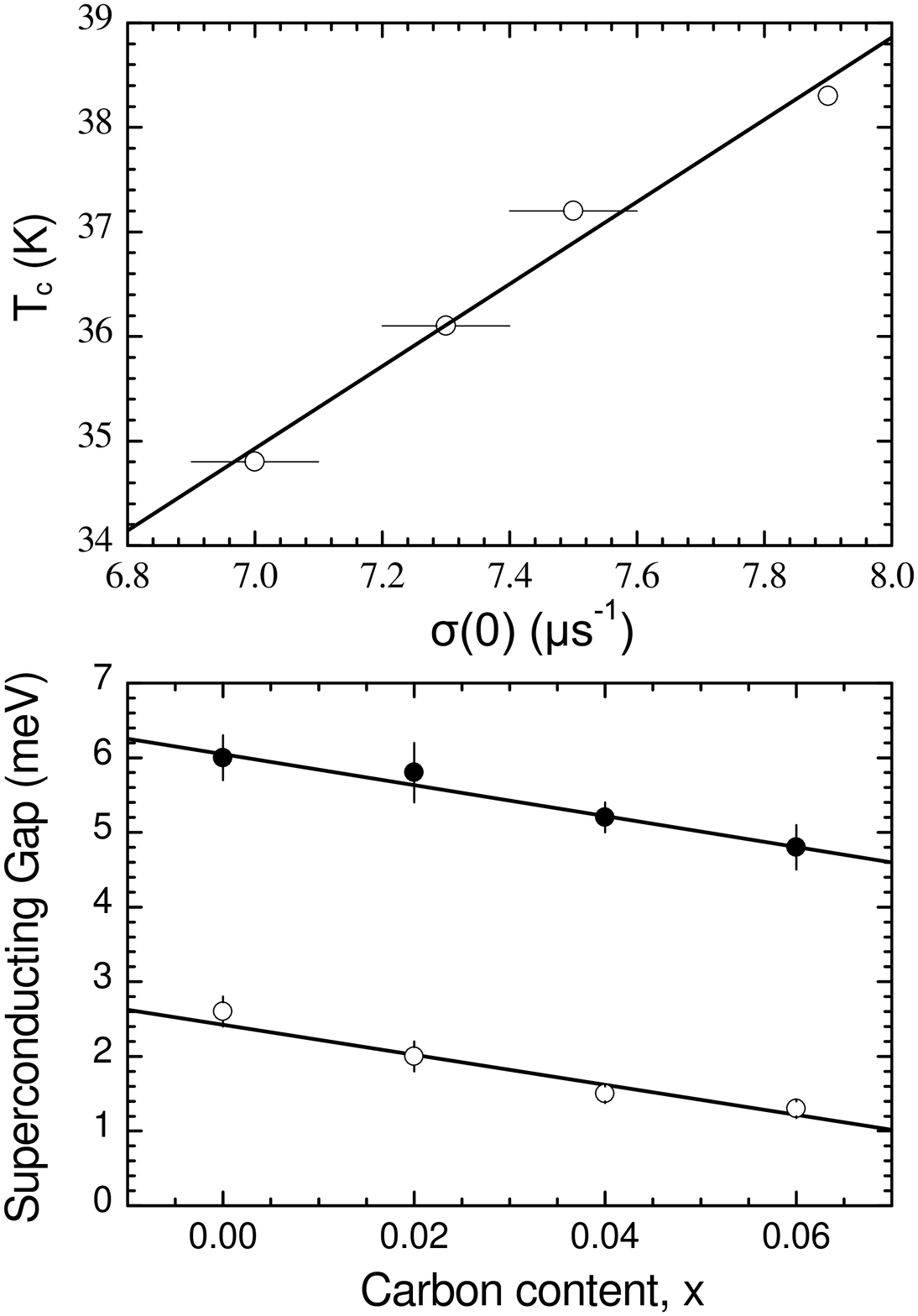}
\caption{Relationship between $T_c$ and low-temperature $\mu^+$ spin depolarisation rate, $\sigma(0)$ 
(upper panel). Evolution of the 
superconducting gap sizes, $\Delta_1$ and $\Delta_2$ at $T$= 0 K with carbon content, $x$ (lower panel). 
The closed symbols correspond to the larger gap, $\Delta_1$ associated with the 2D $\sigma$ sheets and the open 
ones to the smaller gap, $\Delta_2$ associated with the 3D $\pi$ sheets.}
\label{f.2}
\end{figure}

The fitted parameters are summarised in Table I. For comparison the 
corresponding values for \mgb2~\cite{Niedermayer} are also included. 
The evolution of the depolarisation rate $\sigma(0)$ with $T_c$ and of the superconducting gap sizes,
$\Delta_1$(0) and $\Delta_2$(0) with carbon content, $x$ are shown in
Fig.~\ref{f.2}. $\sigma(0)$ 
decreases monotonically with increasing carbon content and decreasing $T_c$ 
(d$\sigma(0)$/d$T_c$= $-$0.25 $\mu$s$^{-1}$K$^{-1}$). The $\sim$11\% decrease in $\sigma(0)$ at $x$= 0.06 
reflects an increase of $\sim$6\% in the in-plane magnetic penetration depth, $\lambda_{ab}(0)$. 
In addition, as the carbon content increases, the two gap sizes at 0 K
shift to smaller values at nearly the same rate (d$\Delta_i$/d$x$= 
$-$20.4 meV, $i$= 1,2). 
$\gamma_1$/$\gamma$ which is a measure of the partial $N$($\epsilon_F$) for
the band with the larger gap also shows a tendency to increase with increasing electron doping.
The superfluid plasma frequency in the $ab$ plane, $\omega_p^{sf}(=c/\lambda_{ab}(0))$ 
which is related to the charge density of the superfluid condensate at 0 K is also included
in Table I. 

\begin{table}[h]
\small
\caption{Extracted parameters for \mgbcx\ ($x$= 0.02, 0.04, 0.06). $\sigma(0)$ is the low-temperature 
$\mu^+$ spin depolarisation rate, $\lambda_{ab}(0)$ is the in-plane magnetic penetration depth, $\Delta_1$ and $\Delta_2$ 
are the superconducting gaps at 0 K, $\gamma$ ($\gamma_1$)
is the (partial) Sommerfeld constant, $\omega_p^{sf}$ is the superfluid
plasma frequency, $n_s$ is the superconducting carrier density and $m_{ab}^*$/$m_e$ is the effective
mass enhancement of the B layers.
The data for \mgb2\ are from Ref.~\cite{Niedermayer} and the $T_c$ values 
from Ref.~\cite{Takenobu}.}
\vspace{3mm}
\begin{tabular}{|c|c|c|c|c|c|c|c|c|}\hline
$x$ & $T_c$ & $\sigma (0)$ & $\lambda_{ab}(0)$& $\Delta_1$ & $\Delta_2$& $\gamma_1$/$\gamma$ & $\omega_p^{sf}$& $n_s$($m_{ab}^*$/$m_e$)$^{-1}$ \\
  & (K) & ($\mu$s$^{-1}$) & (nm) & (meV) & (meV)&  & (eV)& ($\times$10$^{21}$ cm$^{-3}$)\\
\hline\hline
0 & 38.3& 7.9 & 94.7 & 6.0(0) & 2.6(2) & 0.6(2) & 2.08(1) & 3.15 \\
0.02 & 37.2& 7.5(1) & 97.2(6) & 5.8(4) & 2.0(2) & 0.7(2) & 2.03(1)& 2.99(4)\\
0.04 & 36.1& 7.3(1) & 98.5(7) & 5.2(2) & 1.5(1) & 0.8(1) & 2.00(1)& 2.91(4)\\
0.06 & 34.8& 7.0(1) & 100.6(7) & 4.8(3) & 1.3(1) & 0.8(2) & 1.96(1)& 2.79(3)\\
\hline
\multicolumn{9}{l}{ \footnotesize}
\end{tabular}
\end{table}

In the London model at the clean limit, the magnetic penetration depth, $\lambda$ is related to the superconducting 
carrier density, $n_s$ and the effective mass, $m^*$ by 
\begin{equation}
\label{e.4}
\sigma\propto \frac{1}{\lambda^2}=\frac{4 \pi n_s e^2}{m^*c^2}(\frac{1}{1+\xi/l})
\end{equation}
where $\xi$ is the coherence length and $l$ the mean free path. By assuming that \mgbcx\ 
is in the clean-limit ($\xi$= 50 \AA, $l>$ 1000 \AA), the term in brackets is close to unity 
allowing us to estimate the ratio, $n_s/m^*$. 
The calculated values of $n_s/(m_{ab}^*/m_e)$ (where $(m_{ab}^*/m_e)$ is the effective mass 
enhancement in the B layers in units of $m_e$) for \mgbcx\ (Table I) are larger than those obtained 
for high-$T_c$ superconductors and decrease quasilinearly with increasing doping.
In addition, $n_s$ can be converted first into a 2D carrier density, $n_{s2D}$ on the B planes by multiplying 
by the interlayer distance, $c_0$ and then into an effective Fermi temperature, $T_F$
through the expression, $k_B T_F= \hbar^2\pi(n_{s2D}/m^*)$. $T_F$ decreases from $\sim$3200 K in \mgb2\
to $\sim$2700 K in \mgbc6\ resulting in $T_c/T_F\sim$ 0.01. Uemura~\cite{Yemura} has established that 
in contrast to the conventional BCS superconductors in which $T_c/T_F\ll$0.01, the cuprates, fullerides and other exotic 
superconductors have rather high $T_c$ with respect to $T_F$ with $T_c/T_F\sim$ 0.01-0.1.
It is noticeable that the \mgb2\ superconductors also share this feature
and are located in the Uemura plot
between conventional $s$-wave and unconventional 
high-$T_c$ materials.

The electronic band structure of \mgb2 has been extensively 
investigated~\cite{Kortus,An,Medvedeva}. The valence band of \mgb2\ is made up predominantly of B 2$p$ states, which 
form two distinct sets of bands of $\sigma(p_{x,y})$ and $\pi(p_z)$ type whose $k$ dependence differs considerably. 
The most pronounced dispersion for the B $p_{x,y}$ states is along the $\Gamma$-K 
direction of the Brillouin zone (BZ), while a flat zone is formed in the $k_z$ direction ($\Gamma$-A), 
reflecting the 2D character of the boron lattice. These $\sigma$ bands are partially unoccupied creating a 
hole-type conduction band which gives rise to two 2D light-hole and heavy-hole sheets forming coaxial cylinders 
along the $\Gamma$-A BZ direction. The strong coupling of these holes to the optical bond stretching modes 
drives the superconductivity in \mgb2~\cite{Kong}. In addition, B $\pi$ states dominate at the bottom of the 
conduction band, while the Fermi surface associated with these bands consists of a 3D tubular network. 
These bands exhibit maximum dispersion along the $\Gamma$-A direction of the BZ. Theoretical calculations 
estimate that 44\% of $N$($\epsilon _F$) comes from the 2D $\sigma$ cylindrical sheets and the rest 
from the 3D $\pi$ sheets~\cite{Choi}. For the superconducting state of \mgb2, 
the gap is nonzero everywhere on the Fermi surface 
and the gap values are grouped 
in two distinct regions~\cite{Choi}. The larger superconducting gap, $\Delta_1$ on the 2D $\sigma$ cylindrical sheets 
has an average value of $\sim$6.8 meV, while the smaller one, $\Delta_2$ associated with the 3D 
$\pi$-sheets an average value of $\sim$1.8 meV.

Although the electronic structure and the superconducting behaviour of \mgb2\ has been studied theoretically 
in detail, there is no systematic investigation concerning the effect of doping. In the framework of the rigid band model, 
it is expected~\cite{Medvedeva} that doping of the B sublattice with C should lead to a shift of the Fermi level 
to higher energies in the region of the DOS minimum (pseudogap). A rigid band model estimate gives a value of 
$\sim$0.16 electrons/cell in order to fill the $\sigma$ bands~\cite{An}. Electrons are also added to the 
$\pi$ bands that lie in the same energy range, leading to a total doping level $\sim$0.25 electrons/cell. 
Thus it is expected that electron 
doping should lead to both superconducting gaps decreasing gradually. 
However, the differences in the slope $-$dln$T_c$/d$V$ between high-pressure experiments ($\sim$0.28 \AA$^{-3}$) 
on \mgb2~\cite{Prassides} and chemical substitution~\cite{Takenobu} ($\sim$0.37 \AA$^{-3}$), imply that the 
effect of carbon doping is more complicated due to the B layer contraction and theoretical studies beyond the rigid 
band approximation are necessary.

Carbon substitution should also lead to an increase in the interband impurity scattering, which 
should result in the size of the $\sigma$- and $\pi$-gaps converging
to the same value. However, 
theoretical calculations~\cite{Mazin} have shown that the particular electronic structure of \mgb2\ results 
in extremely weak $\sigma\pi$ impurity scattering, even for low quality samples and in
the presence of Mg vacancies, Mg-substitutional impurities and B-site substitutions by N or C. 
The dominant mechanism for impurity scattering is due to intraband scattering of the $\sigma$ and $\pi$ bands 
with the scattering rate inside the $\pi$ bands greater than that of the $\sigma$ bands. 
Intraband scattering does not change $T_c$ and the gap values but influences the penetration 
depth (Anderson's theorem)~\cite{Golubov}. Our experimental data can be described well with a two-gap model 
in which impurity scattering is essentially ignored. This indicates that interband impurity scattering is 
relatively weak and both superconducting gaps are preserved, at least up to $x$= 0.06. Carbon 
doping also reduces both superconducting gaps with this effect relatively more pronounced for the smaller 
gap, $\Delta_2$ associated with the 3D sheets. The error 
in $\gamma_1$/$\gamma_2$ ($\sim$$N_1$($\epsilon_F$)/$N_2$($\epsilon_F$))
is quite large and 
hence it is difficult to extract reliably its dependence on doping. Nonetheless, 
there is a tendency for $\gamma_1$/$\gamma_2$ to increase with increasing doping level. 
In any case, systematic theoretical studies are necessary to shed more light on this issue.

\section{Conclusions}
In conclusion, analysis of the temperature dependence of the TF-\musr\ depolarisation rate for 
\mgbcx\ ($x$= 0.02, 0.04, 0.06) shows that increasing electron doping is accompanied by a decrease of the
low-temperature $\sigma$(0) and the spectral weight, $n_{s}/m^*$, consistent with the decreasing 
density-of-states at the Fermi level. Within a two-gap model, both superconducting gap sizes decrease
almost linearly with the smaller gap affected more on doping, while the interband scattering remains 
relatively weak up to at least $x$= 0.06. 

\acknowledgments
We thank C. Niedermayer and I. I. Mazin for helpful discussions, PSI for provision of beamtime and
A. Amato, K. Brigatti, A. Lappas and I. Margiolaki for help with the experiments. We acknowledge support from the 
Royal Society 
and the Marie Curie Fellowship programme of the EU "Improving the Human Research Potential" under contract 
numbers HPMF-CT-2001-01435 (K. Papagelis) and -01436 (J. Arvanitidis).

\end{document}